\documentclass[twocolumn,showpacs,preprintnumbers,amsmath,amssymb,
superscriptaddress]{revtex4}

\usepackage{graphicx}
\usepackage{dcolumn}
\usepackage{bm}

\usepackage{amsmath}
\usepackage{yhmath}
\usepackage{amssymb}

\date{\today}

\tighten

\newcommand{\disregard}[1]{}

\newcommand{\be}{\begin{equation}}
\newcommand{\ee}{\end{equation}}
\newcommand{\bn}{\begin{eqnarray}}
\newcommand{\en}{\end{eqnarray}}
\newcommand{\ba}{\begin{array}}
\newcommand{\ea}{\end{array}}
\newcommand{\bc}{\begin{center}}
\newcommand{\ec}{\end{center}}
\newcommand{\bml}{\begin{mathletters}}
\newcommand{\eml}{\end{mathletters}}

\newcommand{\shf}{{{\sc shf}}}
\newcommand{\sfo}{{{\sc sf}}}

\newcommand{\ld}{{{\sc ld}}}
\newcommand{\nse}{{{\sc nse}}}
\newcommand{\inm}{{{\sc inm}}}

\tighten

\begin{document}

\draft

\preprint{\fbox{\sc version of \today}}

\title{Global properties of the Skyrme-force-induced nuclear symmetry
energy}

\author{Wojciech Satu{\l}a}
\affiliation{Institute of Theoretical Physics, University of Warsaw,\\
             ul. Ho{\.z}a 69, PL-00 681 Warsaw, Poland}
\affiliation{          KTH (Royal Institute of Technology),\\
           AlbaNova University Center, 106 91 Stockholm, Sweden}

\author{Ramon A. Wyss}
\affiliation{          KTH (Royal Institute of Technology),\\
           AlbaNova University Center, 106 91 Stockholm, Sweden}

\author{Micha{\l} Rafalski}
\affiliation{Institute of Theoretical Physics, University of Warsaw,\\
             ul. Ho{\.z}a 69, PL-00 681 Warsaw, Poland}

\date{\today}

\begin{abstract}
Large scale calculations are
performed to establish the global mass dependence of
the nuclear symmetry energy, $a_{sym}(A)$,
which in turn depends on two basic ingredients:
the mean-level spacing,
$\varepsilon(A)$, and the effective strength of the isovector
mean-potential, $\kappa(A)$.
Surprisingly, our results
reveal that in modern parameterizations including
SLy4, SkO, SkXc, and SkP these two basic ingredients of $a_{sym}$
are almost equal after rescaling them linearly by the isoscalar and the
isovector effective masses, respectively.
This result points toward a new fundamental property of the nuclear
interaction that remains to be resolved. In addition, our analysis
determines the ratio of the surface-to-volume
contributions to $a_{sym}$ to be $\sim$1.6,
consistent with hydrodynamical estimates for the
static dipole polarizability as well as the neutron-skin.

\end{abstract}


\maketitle

The knowledge of the nuclear equation of state ({\sc eos}) for neutron-rich
systems is of fundamental importance for nuclear physics and nuclear
astrophysics. The stability of neutron-rich nuclei, the $r$-process
nucleosynthesis, the structure of neutron star, and the simulations of
supernovae-collapse depend sensitively on the {\sc eos} and, in particular, on
the nuclear symmetry energy ({\sc nse}).
Our study gives new insight into the two basic
ingredients of the symmetry energy, the mean-level spacing,
$\varepsilon(A)$, and the effective strength of the isovector
mean-potential, $\kappa(A)$, revealing the striking property
$\frac{m_0^\star}{m}\varepsilon (A) \approx
\frac{m_1^\star}{m} \kappa (A) $ that holds for
velocity-dependent interactions having
isoscalar $m_{0}^\star$ and isovector $m_{1}^\star$ effective
masses, respectively.
This apparently fundamental property of the effective nuclear
force remains to be explained.

In asymmetric
infinite nuclear matter ({\sc inm}), in the
vicinity of the saturation density, $\rho_0$, the {\sc eos}
(the energy-density per particle) is conveniently
parametrized using the following Taylor expansion:
\bn\label{eos}
\frac{{\cal E}_I(\rho)}{A}  \approx
 -  a_V  & + &  \frac{K_\infty}{18\rho_0^2} (\rho -\rho_0)^2 +  \ldots + \\
   \left[ a_{sym}  + \frac{p}{\rho_0^2}
  (\rho -\rho_0) \right. &   +  & \left.
 \frac{\Delta K_\infty}{18\rho_0^2}  (\rho -\rho_0)^2
 + \ldots \right]  I^2  + \ldots \nonumber
\en
where $I\equiv |N-Z|/A$. The isoscalar~\inm~saturation
density $\rho_0$, and the values of the volume
binding energy $a_V$, the incompressibility parameter $K_\infty$,
and the asymmetry energy $a_{sym}$ serve as primary constraints
for microscopic nuclear models.  For modern Skyrme force parameterizations
which are subject of the present work  $\rho_0\approx 0.16$\,fm$^{-3}$,
$a_V\approx -15.9\pm 0.2$\,MeV, $a_{sym}\approx 32\pm 2$\,MeV, and
$K_\infty\approx 225\pm 25$\,MeV. Higher-order curvature corrections
to the~\nse, $p$, $\Delta K_\infty$, are rather poorly constrained.
All these values are derived from the studies of finite nuclei.

Integrating out the
$\boldsymbol r$-dependence from the energy-density
leads then to the semi-empirical mass formula ({\sc ld}) for
the energy per particle which is conventionally written as:
\be\label{ldrop}
\frac{E}{A} = -a_V + \frac{a_S}{A^{1/3}}+ 
 \left[
a_{sym}^{(V)} - \frac{a_{sym}^{(S)}}{A^{1/3}} +\ldots \right]
\left( I^2+ \lambda\frac{I}{A} \right) + \ldots ,
\ee
where $a_S$ and  $a_{sym}^{(S)}$ are coefficients defining
contributions from the
surface energy and the surface part of the symmetry energy, respectively.
There is at present no consensus concerning the magnitude, $\lambda$,
as well as origin of the term linear in $\sim$$I$, which is often called the
Wigner energy.

Another controversy exists concerning the surface contribution
 to the~\nse. The values of the surface-to-volume ratio $r_{S/V} =
 a_{sym}^{(S)} / a_{sym}^{(V)}$ quoted in the literature vary
 strongly. For example,
 Danielewicz~\cite{[Dan03]} estimates it to be
 $ 2.0 \leq r_{S/V} \leq 2.8$, the mass formula
 of Ref.~\cite{[Mol95]} yields $r_{S/V} \approx 1.6 $ while
 the hydrodynamical-type models that include properly
 polarization of the isovector density predict
 $r_{S/V} \approx 2 $~\cite{[Lip82]} which,
 according to Ref.~\cite{[Boh81]}, is
 consistent within $10\div 20$\%
 with Skyrme-force ({\sc sf}) calculations.

The main objective  of this work is to study the symmetry energy
within the Skyrme-Hartree-Fock (\shf)~model.
Notwithstanding, that a deeper understanding of the
symmetry energy is crucial in order to reach
a consensus on the existing variety of~\sfo~parameterizations,
or to constrain the coupling constants of a more general
local energy density functional ({\sc ledf}).
Our results point toward a deeper relation between the
average level spacing and the strength of the
mean isovector potential which has not been addressed hitherto.
The calculations presented below also
allow us to determine the surface-to-volume ratio of
the~\shf~symmetry energy.

In our previous
letter~\cite{[Sat03]} we have demonstrated that the~\shf~symmetry
energy behaves rather unexpectedly according to the formula:
\be\label{esym}
   E_{sym}^{(SHF)} = \frac{1}{2} \varepsilon (A,T_z) T^2 +
                     \frac{1}{2} \kappa (A,T_z)  T(T+1) \, ,
\ee
where $\varepsilon (A,T_z)\approx \varepsilon(A) $ and
$\kappa (A,T_z) \approx \kappa (A)$ are fairly independent on
$T_z$, at least for $T_z \geq 8$, and
denote the mean-level
spacing at the Fermi energy in iso-symmetric nuclei and
effective strength of the isovector
mean-potential emerging within the~\shf, respectively.
More precisely, $\kappa$ is related to the isovector
part of the~\sfo~induced {\sc ledf} ({\sc s-ledf})
${\mathcal H}({\boldsymbol r}) = \sum_{t=0,1}
{\mathcal H}_{t} ({\boldsymbol r})$:
\begin{equation}\label{teven} {\mathcal H}_{t}
 = C_{t}^{\rho} \rho_t^2 + C_{t}^{\Delta\rho}
\rho_t\Delta\rho_t + C_t^\tau \rho_t\tau_t + C_t^J \overleftrightarrow{J}_t^2
+ C_t^{\nabla J} \rho_t {\boldsymbol \nabla}\cdot {\boldsymbol J}_t.
\end{equation}
Definitions of all local
densities and currents $\rho, \tau, \overleftrightarrow{J}$  as well as the
explicit expressions for coupling constants $C_t$ can be found
in numerous references and we follow the notation used in~Ref.~\cite{[Dob00]}.
Due to the isoscalar-density dependence
of the~\sfo~, the coupling constants $C_t^\rho[\rho_0]$ of
the {\sc s-ledf} are functionals of $\rho_0$, giving rise to the
isoscalar rearrangement mean-potential $U_0 = \sum_{t=0,1}
\frac{\partial C_t^\rho}{\partial \rho_0} \rho_t^2$.  Since our procedure
of extracting $\varepsilon$ and $\kappa$
involves setting the $C_1\equiv 0$, see~Ref.~\cite{[Sat03]},
part of the $U_0$ related to the $C_1^\rho$
is formally treated as being related to the isovector degrees of freedom.
Note that this separation is 
is consistent with the way the symmetry energy
        constraint is superimposed on the ~\sfo~.
It does not affect
the generality of our approach and
similar analysis can be performed also within the much wider
class of the Hohenberg-Kohn-Sham {\sc ledf} theories.

In the present work we focus on the global mass dependence of
the~\shf~values of $\varepsilon (A)$ and $\kappa(A)$ and
perform a systematic calculation covering all
even-even nuclei having $20\leq A\leq 128$ from $N=Z$ to
almost the neutron drip line.
Coulomb and pairing effects are
disregarded i.e. the emphasis is on the strong interaction
acting in the particle-hole channel.
The calculations are performed for a set of
different {\sc sf} parameterizations as the SkP~\cite{[Dob84]},
SkXc~\cite{[Bro98]}, Sly4~\cite{[Cha97]}, SkO~\cite{[Rei99]},
SkM$^\star$~\cite{[Bar82]}, and SIII~\cite{[Bei75]},
using the \shf~code
{\sc hfodd} of Dobaczewski {\it et al.\/}~\cite{[Dob00],[Dob04]}.

The procedure used to extract $\varepsilon (A,T_z)$ and
$\kappa (A,T_z)$ follows exactly the one outlined in Ref.~\cite{[Sat03]}.
First, we set all the isovector coupling constants $C_1\equiv 0$ in the
{\sc s-ledf} (\ref{teven}) and extract $\varepsilon (A,T_z)$ by
comparing calculated excitation energy
$\Delta E_{SHF}^{(t=0)} (A,T_z)\equiv  E_{SHF}^{(t=0)}(A,T_z) -
   E_{SHF}^{(t=0)} (A,0) $ to:
\be
   \Delta E_{SHF}^{(t=0)}(A,T_z)
   = \frac{1}{2}\varepsilon(A,T_z) T^2.
\ee
In the next step, we compute the total~\shf~binding energy
$E_{SHF}(A,T_z)$ and compare:
\be\label{kap}
   \Delta E_{SHF}(A,T_z) - \Delta E_{SHF}^{(t=0)}(A,T_z)
   = \frac{1}{2}\kappa (A,T_z) T(T+1),
\ee
in order to determine $\kappa(A,T_z)$.

\begin{figure}[t]
\includegraphics[width=8.0cm,clip]{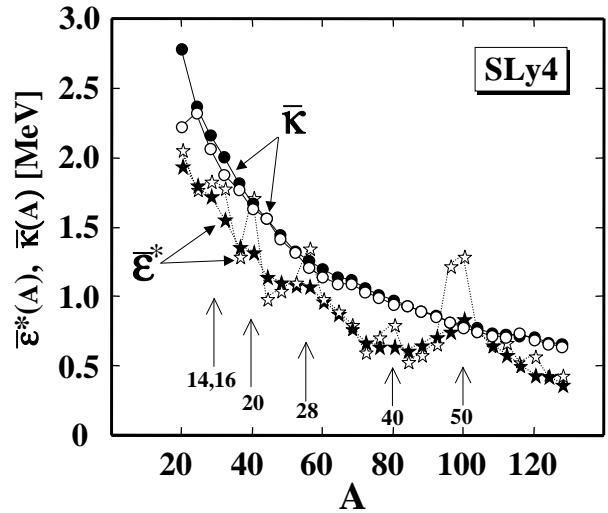}
\caption[]{The isoscalar effective mass scaled values of
${\bar\varepsilon}^\star (A) \equiv  \frac{m_0^\star}{m} {\bar\varepsilon}(A)$
(stars) and ${\bar\kappa} (A)$ (circles) calculated using the~\shf~method with
SLy4 parametrization. Open symbols denote ${\bar\varepsilon}^\star (A)$ and
${\bar\kappa}(A)$ averaged over all the calculated nuclei. Filled symbols mark
smoothed values of ${\bar\varepsilon}^\star (A)$ and ${\bar\kappa} (A)$
calculated using a restricted set of data. Vertical arrows indicate major shell
gaps. Note the strong influence of shell structure on ${\bar\varepsilon}^\star
(A)$ and the smooth behavior of ${\bar\kappa}(A)$. } \label{sl4}
\end{figure}

For each $A$ and small $T_z$, the values
of $\varepsilon (A,T_z)$ oscillate quite rapidly.
However, they clearly tend to stabilize for $T_z\geq 8$ where
$\varepsilon (A,T_z)\approx \varepsilon (A)$.
The values of $\kappa (A,T_z)$ appear to stabilize faster
and $\kappa (A,T_z)\approx\kappa (A)$ essentially already
for $T_z \geq 4$. It should be mentioned that in the case of
the SkO parameterization the formula~{(\ref{kap}) does work
only approximately. For this force we observe an
enhancement in the linear, $\sim$$T$.
This effect is, however, much weaker than the analogous
effect found recently within relativistic mean field~\cite{[Ban05]}
where it restores the $E_{sym}\sim T(T+1)$ dependence of the
total~\nse.

For further quantitative analysis of
the mass dependence of the~\nse~we use the mean values of
${\bar{\varepsilon}}(A)$ and ${\bar{\kappa}} (A)$.
These averages over $T_z$ at fixed
$A$ are calculated using the following restricted set of
nuclei: $T_z\geq 4$ for $A =20$;  $T_z\geq 6$ for $A=24$; and $T_z\geq 8$ for
$A\geq 28$. By using a restricted set of nuclei we smooth out
both  $\bar{\varepsilon}(A)$ and $\bar{\kappa} (A)$ curves
in order to diminish the possible influence of shell structure.

The global mass dependence
of the two components of the symmetry energy, $\bar\varepsilon$
and $\bar\kappa$ are presented in Fig.~\ref{sl4}. Although representative
for the SLy4 parametrization,
the figure shows several
features which appears to be independent
of the type of the {\sc sf} parametrization. These
universal features include:
({\it i\/}) strong dependence of $\bar{\varepsilon} (A)$ on
kinematics (shell effects); ({\it ii\/}) almost no
dependence of $\bar{\kappa} (A)$ on kinematics; ({\it iii\/})
clear surface ($\sim \frac{1}{A^{4/3}}$) dependence reducing
the dominant volume term ($\sim \frac{1}{A}$) in both
${\bar \varepsilon} (A)$ and  ${\bar \kappa} (A)$.

Indeed, the values of  $\bar{\varepsilon} (A)$  show characteristic
kinks close to double-(semi)magic $A$-numbers. These kinks
are magnified when all the calculated nuclei are used (no smoothing)
to compute $\bar{\varepsilon} (A)$,
but without affecting qualitatively the overall profile of
the curve. On the other hand, $\bar{\kappa}(A)$ is almost perfectly smooth
with barely visible traces of shell structure.
It confirms our earlier conclusion~\cite{[Sat03]}
that the gross features of the Skyrme
isovector mean potential can be almost
perfectly quantified by a smooth curve parametrized by
a small number of global parameters.

\begin{table*}
\begin{center}
\begin{tabular}{lrrrrrrrrrrrrrrrrrrr}
\hline
\hline
     &  $m_0^*/m$     &  $m_1^*/m$  & $\quad$ &
$\varepsilon^\star_V$  &  $\varepsilon^\star_S$  & $r_\varepsilon$  & $\quad$ &
$\kappa^\star_V$  &  $\kappa^\star_S$ & $r_\kappa$ &$\quad$ &  $r^\star_V$
      &  $r^\star_S$ &  $r^\star_{(\infty)}$ &
      $\quad$ & $a_{sym}^{(\infty)}$  &  $a_{sym}^{(V)}$ & $a_{sym}^{(S)}$
      &  $r_{S/V}$  \\
\hline
\hline
\mbox{SLy4}  & 0.695 & 0.800 & &  94.5 & 147.5 & 1.56 & &
                                  94.7 & 137.5 & 1.45 & &
                                  1.00 & 1.07  & 1.07 & &
                                  32.0 & 31.8  & 48.0 & 1.51
                                              \\
\hline
\mbox{SkXc}  & 1.006 & 0.752 & & 108.6 & 164.3 & 1.51 & &
                                 107.6 & 165.2 & 1.54 & &
                                 1.01 & 0.99   & 0.88 & &
                                 30.1 & 31.4   & 47.9 & 1.53
                                              \\
\hline
\mbox{SkP}   & 1.000 & 0.741 & & 108.8 & 175.1 & 1.61 & &
                                 106.0 & 163.1 & 1.54 & &
                                 1.03 & 1.07   & 0.95 & &
                                 30.0 & 31.5   & 49.4 & 1.57
                                              \\
\hline
\mbox{SkO}   & 0.896 & 0.852 & & 107.2 & 166.2 & 1.55 & &
                                 110.6 & 176.1 & 1.59 & &
                                  0.97 & 0.94  & 0.79 & &
                                  32.0 & 31.2  & 49.0 & 1.57
                                              \\
\hline
\mbox{SkM*}  & 0.789 & 0.653 & & 106.3 & 180.9 & 1.70 & &
                                  71.4 & 107.3 & 1.50 & &
                                  1.49 & 1.69  & 1.37 & &
                                  30.0 & 30.5  & 49.2 & 1.61
                                              \\
\hline
\mbox{SIII}  & 0.763 & 0.655 & &  97.5 & 143.8 & 1.47 & &
                                  75.2 & 103.2 & 1.37 & &
                                  1.30 & 1.39  & 1.34 & &
                                  28.2 & 30.3  & 43.3 & 1.43
                                              \\

\hline
\hline
\end{tabular}
\caption[A]{The table
includes: the isoscalar, $m_0^\star /m$, and the isovector, $m_1^\star/m$,
effective masses;
the volume, $\varepsilon^\star_V\, (\kappa^\star_V)$,
the surface $\varepsilon^\star_S\, (\kappa^\star_S)$,
and the ratios $r_\varepsilon = \varepsilon^\star_S/\varepsilon^\star_V
\, (r_\kappa = \kappa^\star_S/\kappa^\star_V)$
of the expansion coefficients~(\ref{v-vs-s})
of the isoscalar effective mass scaled ${\bar \varepsilon}^\star (A)$;
and the isovector effective mass scaled
${\bar \kappa}^\star (A) \equiv \frac{m_1^\star}{m} {\bar \kappa}(A)$,
respectively;
the ratios of volume $r^\star_V = \varepsilon^\star_V/\kappa^\star_V$
and surface  $r^\star_S = \varepsilon^\star_S/\kappa^\star_S$
expansion coefficients
and the~\inm~estimate $r^\star_{(\infty)} = \varepsilon^\star_{(\infty)}
/\kappa^\star_{(\infty)}$; the~\inm~estimate
$a_{sym}^{(\infty)}$ and the calculated values $a_{sym}^{(V)}$,
$a_{sym}^{(S)}$ and $r_{S/V} = a_{sym}^{(S)}/a_{sym}^{(V)}$ of
the symmetry energy coefficient as defined in~(\ref{ldrop}).
The values of  $\varepsilon^\star_{V(S)}$,   $\kappa^\star_{V(S)}$,
$a_{sym}^{(\infty)}$, $a_{sym}^{(V)}$, and  $a_{sym}^{(S)}$ are given in
MeV. }\label{wyniki}
\end{center} \end{table*}

In the analysis of a leptodermous expansion of
$\bar{\varepsilon}(A)$ and $\bar{\kappa}(A)$ we consider
volume ($V$) and surface ($S$) terms:
\be\label{v-vs-s}
   \bar{\varepsilon}(A) = \frac{\varepsilon_V}{A} -
                    \frac{\varepsilon_S}{A^{4/3}}, \quad \quad
   \bar{\kappa}(A) = \frac{\kappa_V}{A} -
                    \frac{\kappa_S}{A^{4/3}}.
\ee
The values of the isoscalar-effective-mass-scaled
expansion coefficients $\varepsilon_V^\star$, $\varepsilon_S^\star$
as well as the values of the isovector-effective-mass-scaled
expansion coefficients $\kappa_V^\star$, $\kappa_S^\star$
are collected in Tab.~\ref{wyniki}. First of all,
let us observe that the calculated value of $\varepsilon_V^\star\approx
100$MeV corresponds to the pure Fermi gas estimate
$\varepsilon_{FG}$. This result can be understood based on
the analytical expression for the Skyrme force~\nse~coefficient
in the limit of symmetric~\inm , $a_{sym}^{(\infty)}$, provided that
the standard textbook formula is rewritten in the following way:
\begin{eqnarray}\label{nmsym}
a_{sym}^{(\infty)} & = &\frac{1}{8}\varepsilon_{FG} \left( \frac{m}{m_0^\star}
\right) +
\left[ \left( \frac{3\pi^2}{2}\right)^{2/3} C_1^\tau \rho^{5/3}
+ C_1^\rho \rho \right]\nonumber \\
&\equiv &\frac{1}{8}\left[ \varepsilon_{(\infty)} +\kappa_{(\infty)}
\right],
\end{eqnarray}
where $C_1^\tau$ and $C_1^\rho$ define the isovector
part of the~{\sc s-ledf}, see Eq.~(\ref{teven}).
Eq.~(\ref{nmsym}) clearly separates the contributions
from the isovector and the isoscalar part and relates the latter to
the single particle energies in~\inm,
$\varepsilon_p = \frac{{\boldsymbol p}^2}{2m} + \Sigma({\boldsymbol p},
\varepsilon_p ) = \frac{{\boldsymbol p}^2}{2m_0^\star}$,
with a self-energy term,  $\Sigma({\boldsymbol p},
\varepsilon_p )$, that describes  the interaction with the nuclear medium
incorporated into the isoscalar effective mass. Hence, Eq.~(\ref{nmsym})
further supports our interpretation of the~\nse~strength.

The most striking result of our
analysis is the {\bf near-equality}
of $\bar{\varepsilon}^\star \approx \bar{\kappa}^\star$
occurring for all modern parameterizations,
see Tab.~\ref{wyniki} and Fig.~\ref{e-k-star}.  Indeed,
$\bar{\varepsilon}^\star$ differs from ${\bar\kappa}^\star$
only for old parameterizations like the SIII and SkM$^*$.
This result confirms the rather loose claims often
appearing in textbooks that "{\it the kinetic energy\/}
[$\varepsilon_{FG}$] {\it and the isovector mean-potential contribute to the
$a_{sym}$ in a similar way\/}" is indeed correct but only after disregarding
non-local effects.
To our knowledge, it has never been discussed why this apparently independent
quantities should be similar.

\begin{figure}
    \centering
\includegraphics[width=8.0cm,clip]{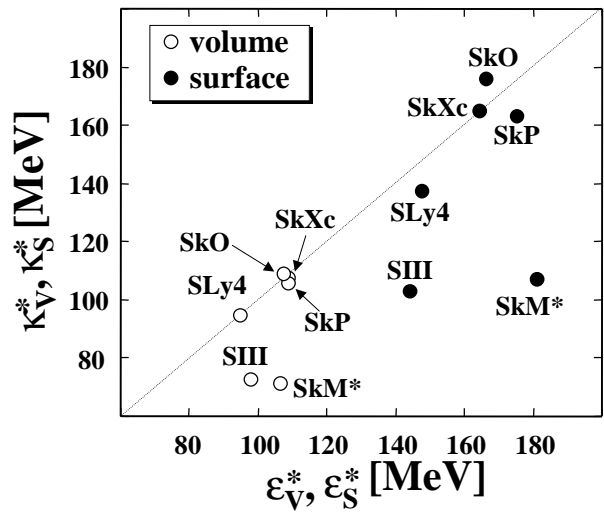}
\caption[]{The correlation between the effective
mass scaled volume
$\varepsilon_V^\star$ and $\kappa_V^\star$ (open dots) and the surface
$\varepsilon_S^\star$ and $\kappa_S^\star$ expansion coefficients.
Note, that except for the SIII and SkM$^\star$ interactions, the expansion
coefficients are equal, see also Tab.~\ref{wyniki}. } \label{e-k-star}
\end{figure}

The relation $\bar\varepsilon^\star \approx \bar\kappa^\star$
could be accidental. However, given
the fact that these modern Skyrme forces
have been fitted in a rather different manner,
suggests that the near equality is of fundamental nature.
To find an explanation of this result will be
a challenge for further studies.
Note that the analytical relation (\ref{nmsym}) does not show
any explicit scaling due to the presence of the
isovector effective mass, $m_1^\star$. This
is probably not very surprising since $m_1^\star$
defines the enhancement in the energy weighted
sum rule for the translational symmetry violating (finite nucleus)
dipole mode. Moreover, the key isovector coupling constants
$C_1^\tau$ and  $C_1^\rho$ vary completely erratically from force
to force, although we have found that they are linearly correlated for all
the~\sfo~studied here.
It indicates that they are neither well
understood  nor well constrained.

The advantage of our interpretation of $\varepsilon (A)$
becomes evident when evaluating
$r_\varepsilon$
using the semi-classical approximation~\cite{[Bal70]}
in straight forward fashion.  The appropriate formula
which takes into account the diffuseness of the potential,
see~Ref.~\cite{[Sto85]}:
\be\label{eval}
   \varepsilon (A) \sim g(\epsilon_F)^{-1} \sim
   \frac{1}{A}\left( 1 -  \frac{\pi}{4k_F} \frac{S_M}{V_M}
   + \ldots \right),
\ee
where $ g(\epsilon_F)$ is the level density at the Fermi energy,
$k_F  \approx 1.36$\,fm$^{-1}$ while $V_M$  and
$S_M$ denote volume and surface
matter-distribution, respectively. Assuming spherical geometry
$\frac{S_M}{V_M} \approx
\frac{3}{r_o A^{1/3}}$ and adopting for $r_o\approx 1.14$\,fm, i.e. the value
consistent with the standard
Skyrme force saturation density $\rho_0\approx 0.16$\,fm$^{-1}$,
one obtains $r_\varepsilon \approx \frac{3\pi}{4k_F r_o}\approx
1.52$ which is indeed very close to the calculated ratios, see
Tab.~\ref{wyniki}.

The~\shf~models yield $r_{S/V}\sim 1.6$ in accordance with
the~\ld~ratio~\cite{[Mol95]}.
The static dipole polarizability (SDP) $\alpha_D$
[$\sigma (\omega)$ denotes photo-absorption cross-section]:
\begin{equation}\label{sigma2}
\sigma_{-2} \equiv  \int \frac{\sigma (\omega)}{\omega^2} d\omega \equiv
2\pi^2 \frac{e^2}{\hbar c} \alpha_D,
\end{equation}
provides an independent cross-check of the ratio $r_{S/V}$.
Indeed, using the so called hydrodynamical
model simple estimate for $\alpha_D$  can be derived~\cite{[Lip82]}:
\begin{equation}\label{alphad}
\alpha_D \approx  \alpha_D^{\text{(M)}} \left(1+\frac{5}{3}\frac{r_{S/V}}
{A^{1/3}} + \ldots \right),
\end{equation}
where $\alpha_D^{\text{(M)}} = \frac{1}{24} \frac{\langle r^2
\rangle}{a_{sym}^{(V)}}$
is the so called Migdal SDP
value~\cite{[Mig44]} which is valid for large systems with negligible surface
contribution. Using $\langle r^2\rangle = \frac{3}{5} R^2 A$ where $R =
1.2A^{1/3}$fm  and  $a_{sym}^{(V)}=30$MeV one obtains, in the Migdal's limit,
the following numerical estimate
$\sigma_{-2}^{\text{(M)}}A^{-5/3}\approx 1.73$$\mu$b/MeV. Using this estimate
and the experimental value of
 $\sigma_{-2}^{(exp)}A^{-5/3}\approx 2.7\pm 0.2$$\mu$b/MeV which is almost
constant for $A\geq 100$~\cite{[Ber77],[Boh81]}, one obtains $r_{S/V} \sim
1.65$ what is again
consistent with our estimates based on the~\shf~calculations.

The neutron skin thickness  is another quantity
which sensitively depends on isovector properties. Neutron
rms radii are even today rather poorly known and can therefore not
be used to constrain the {\sc ledf} or the {\sc eos}, see
however~\cite{[Bro00]}. For heavy nuclei one can continue to apply the
hydrodynamical model of Ref.~\cite{[Lip82]} and evaluate the neutron skin
thickness:
\begin{equation}\label{deltar} \frac{\delta r^2}{\langle r^2
\rangle}  \approx \frac{N-Z}{A} \left\{  1 + \frac{2}{3}
\frac{r_{S/V}}{A^{1/3}} - \ldots \right\}.
\end{equation}
Interestingly, Eq.~(\ref{deltar}) does not
depend on the bulk~\nse~coefficient but only on
the ratio $r_{S/V}$. Formula~(\ref{deltar}) taken at $r_{S/V}\sim 1.65$
gives again a very consistent
results with our microscopic~\shf~calculations.

In summary, the global mass dependence of the~\nse~strength $a_{sym}(A)$
and its two basic ingredients related to the mean-level spacing,
$\varepsilon (A)$, and to the mean-isovector potential, $\kappa (A)$
is studied in detail within the~\shf~theory.
Our interpretation of the symmetry energy enables us
to unambiguously establish the
surface-to-volume ratio
of $a_{sym} (A)$, $r_{S/V}\approx 1.6$
in agreement with the~\ld~value of Ref.~\cite{[Mol95]}. 
This ratio is consistent with simple hydrodynamical estimates
for the SDP and neutron skin thickness.
The most striking results of our calculations is
the near-equality
of $\bar{\varepsilon}^* \approx \bar{\kappa}^*$ revealing that {\it
contribution to $a_{sym}$ due to the mean-level spacing and due to
the mean-isovector potential are similar\/} but only after
disregarding non-local effects.
Whether this is a fundamental property of the nuclear mean field
is an open question that requires further studies.

\smallskip

This work has been supported by the G\"oran Gustafsson Foundation,
the Swedish Science Council (VR),
the Polish Committee for Scientific Research (KBN) under
contract 1~P03B~059~27, and the
Foundation for Polish Science (FNP).


\end{document}